\documentclass[12pt]{article}
\def\a{\alpha}
\def\b{\beta}
\def\g{\gamma}
\def\vt{\vartheta}
\def\d{\delta}

\def\cA{{\cal A}}
\def\cF{{\cal F}}
\def\cJ{{\cal J}}

\def\cL{{\cal L}}

\usepackage{amsmath}
\usepackage{amssymb,amsthm,graphicx,citesort}
\usepackage[ansinew]{inputenc}
\usepackage{stmaryrd}
\usepackage{multibox}
\usepackage{graphicx, epic, eepic, color}% Include figure files
\usepackage{dcolumn}% Align table columns on decimal point
\usepackage{bm}% bold math
\usepackage[ansinew]{inputenc}
\usepackage{times}

\begin{document}

\title{An assessment of Evans' unified field
  theory II}

\author{Friedrich W.~Hehl\footnote{Institute for Theoretical Physics,
    University at Cologne, 50923 K\"oln, Germany, and Department of
    Physics and Astronomy, University of Missouri-Columbia, Columbia,
    MO 65211, USA} \; and Yuri N.~Obukhov\footnote{Institute for
  Theoretical Physics, University at Cologne, 50923 K\"oln, Germany,
  and Department of Theoretical Physics, Moscow State University,
  117234 Moscow, Russia}}

\date{08 March 2007, {\it file AssessII06.tex, fwh}}

\maketitle

\begin{abstract}
  Evans developed a classical unified field theory of gravitation and
  electromagnetism on the background of a spacetime obeying a
  Rie\-mann-Cartan geometry. In an accompanying paper I, we analyzed
  this theory and summarized it in nine equations. We now propose a
  variational principle for Evans' theory and show that it yields two
  field equations. The second field equation is algebraic in the
  torsion and we can resolve it with respect to the torsion. It turns
  out that for all physical cases the torsion vanishes and the first
  field equation, together with Evans' unified field theory, collapses
  to an ordinary Einstein equation.
\end{abstract}

\begin{footnotesize}

\end{footnotesize}

\newpage

\section{Introduction}

In an accompanying paper \cite{AssI}, called I in future, we
investigated the unified field theory of Evans
\cite{Evans2003,Evans2005a}. We take the notation and the conventions
{}from I, where also more references to Evans' work can be found. We
assume that the reader is familiar with the main content of part I
before she or he turns her or his attention to the present paper. In I
we were able to reduce Evans' theory to just nine equations, which we
will list again for convenience.

Spacetime obeys in Evans' theory a Riemann-Cartan geometry
(RC-geometry) that can be described by an orthonormal coframe
$\vt^\a$, a metric $g_{\a\b}=\mbox{diag}(+1,-1,$ $-1,-1)$, and a Lorentz
connection $\Gamma^{\a\b}=-\Gamma^{\b\a}$. In terms of these
quantities, we can define torsion and curvature, respectively:
\begin{eqnarray}\label{torsion*}
  T^\a&:=&D\vt^\a\,,\\ \label{curv'}
  R_\a{}^\b&:=&d\Gamma_\a{}^\b-\Gamma_\a{}^\g\wedge\Gamma_\g{}^\b\,.
\end{eqnarray}
The Bianchi identities and their contractions follow
therefrom.

Evans proposes an extended electromagnetic field with the potential
$\cA^\a$. By Evans' ansatz, this potential is postulated to be
proportional to the coframe
\begin{equation}\label{Ansatz}
\cA^\a=a_0\,\vt^\a\,,
\end{equation}
with some constant $a_0$. The electromagnetic field strength is
defined according to
\begin{equation}\label{DefF}
\cF^\a:=D\cA^\a\,.
\end{equation}

The extended homogeneous and inhomogeneous Maxwell equations read in
Lorentz covariant form
\begin{equation}\label{both}
D\cF^\a=R_\b{}^\a\wedge \cA^\b\qquad\mbox{and}\qquad
D^\star\cF^\a={}^\star\!R_\b{}^\a\wedge\cA^\b\,,
\end{equation}
respectively. Alternatively, with Lorentz non-covariant sources and
with partial substitution of (\ref{Ansatz}) and (\ref{DefF}), they can
be rewritten as
\begin{eqnarray}\label{Evanshom'}
  d\cF^\a=\Omega_0\,\cJ^\a_{\rm hom}\,,\quad&&  \cJ^\a_{\rm hom}
:=\frac{a_0}{\Omega_0}\left( R_\b{}^\a\wedge 
\vt^\b-\Gamma_\b{}^\a\wedge T^\b\right) \,,
\\d\,^\star\!\cF^\a=\Omega_0\,\cJ^\a_{\rm inh}\,,\quad&&  \cJ^\a_{\rm inh}
:=\frac{a_0}{\Omega_0}\left( 
{}^\star\! R_\b{}^\a\wedge 
\vt^\b-\Gamma_\b{}^\a\wedge {}^\star\!T^\b\right)\,.\label{Evansinh'}
\end{eqnarray}
 
In the gravitational sector of Evans' theory, the Einstein-Cartan
theory of gravity (EC-theory) was adopted by Evans.
Thus, the field equations are those of Sciama \cite{Sciama1,Sciama2}
and Kibble \cite{Kibble}, which were discovered in 1961:
\begin{eqnarray}\label{Sc1*}
  \frac 12\,\eta_{\a\b\g}\wedge
  R^{\b\g}&\!=\!&\kappa\,\Sigma_\a=\kappa\left(\Sigma_\a^{\rm mat}+\Sigma_\a^{\rm
      elmg} 
  \right)\,,\\ \frac 12\,\eta_{\a\b\g}\wedge T^\g
  &\! =\!&\kappa\,\tau_{\a\b}=\kappa\left(\tau_{\a\b}^{\rm mat}+\tau_{\a\b}^{\rm
      elmg} \right)\,.\label{Sc2*}
\end{eqnarray}
Here $\eta_{\a\b\g}=\,^\star\!\left(\vt_\a\wedge\vt_\b\wedge\vt_\g
\right)$. The total energy-momentum of matter plus electromagnetic
field is denoted by $\Sigma_\a$, the corresponding total spin by
$\tau_{\a\b}$.  

This is the set-up. What we will do here is to propose a new
variational principle that describes Evans' theory. We will derive the
field equations and will discuss their properties.

\section{Closing loopholes in Evans' theory}

It is apparent that there exist a couple of
loopholes in Evans' theory. Apart {}from announcing the second field
equation (\ref{Sc2*}) only verbally and without specifying any
formula, the right-hand sides of the two field equations are left open
in Evans' approach. How is the energy-momentum $\Sigma_\a^{\rm elmg}$
of Evans' field defined, how the spin $\tau_{\a\b}^{\rm elmg}$?
Silence is the only answer in Evans' verbose publications. In order to
have a better grip on Evans' theory, we decided to develop it a bit
further.

{}From the summary of Evans' theory it becomes clear that the {\it
  geometrical} equations (\ref{torsion*}),(\ref{curv'}) and the
{\it gravitational} equations (\ref{Sc1*}),(\ref{Sc2*}) represent the
viable EC-theory of gravitation that is distinct {}from general
relativity by an additional spin-spin contact interaction which only
acts at very high matter densities, see the review \cite{RMP}. If the
sources in the framework of the EC-theory are the {\it Maxwell field}
$A$ (with $F=dA$) and some matter fields $\Psi$, we have the
variational principle
\begin{eqnarray}\label{ECMax}
  L_{\rm EC}&=&-\frac{1}{2\kappa}\,^\star\!\left(\vt_\a\wedge\vt_\b
 \right)\wedge R^{\a\b}-
  \frac{1}{2\Omega_0}\,F_\a\wedge\,^\star\!F^\a\nonumber\\&& +L_{\rm
    mat}\left(\vt^\a, \Psi^{\a\b...},D\Psi^{\a\b...}\right)\,.
\end{eqnarray}
The matter fields $\Psi$ are supposed to be minimally coupled to
gravity and to electromagnetism. Variation with respect to $A$ yields
the inhomogeneous Maxwell equation $d\,^\star\!F= \Omega_0\,J$, with
$J=\d L_{\rm mat}/\d A$, variation with respect to $\vt^\a$ and
$\Gamma^{\a\b}$ the gravitational field equations (\ref{Sc1*}) and
(\ref{Sc2*}), with $\Sigma_\a^{\rm elmg}$ and $\tau_{\a\b}^{\rm elmg}$
substituted by $\Sigma_\a^{\rm Maxw}$ and $\tau_{\a\b}^{\rm Maxw}$,
respectively. This is conventional wisdom. Thereby, we also find the
canonical energy-momentum and the spin angular momentum of the Maxwell
field:
\begin{eqnarray}\label{Maxwell}
  \Sigma_\alpha^{\rm Maxw}&:=& =-\frac{\d L_{\rm Maxw}}{\d \vt^\a}=
\frac{1}{2\Omega_0}\left[F\wedge(e_\alpha\rfloor
    {}^\star\! F ) -  {}^\star\! F\wedge (e_\alpha\rfloor F)\right]\,,\\
  \tau_{\a\b}^{\rm Maxw}&:=&-\frac{\d L_{\rm Maxw}}{\d \Gamma^{\a\b}}=0\,.
\end{eqnarray}

However, in Evans' theory, instead of the Maxwell field, we have
Evans' extended electromagnetic field. Then the questions arise what
the sources on the right-hand-sides of the gravitational field
equations (\ref{Sc1*}) and (\ref{Sc2*}) are and what the extended
electromagnetic field may contribute to them. For the time being, we
forget Evans' ansatz, that is, we develop a field theoretical model
{\em before Evans' ansatz (\ref{Ansatz}) is substituted}.

\subsection{Auxiliary Lagrangian}

We proceed like in Maxwell's theory. We pick Evans' $\cA^\a$ potential
as the electromagnetic field variable and define the field strength
$\cF^a=D\cA^\a$. Then $D\cF^\a=DD\cA^\a=R_\b{}^\a\wedge \cA^\b$ is the
homogeneous equation (\ref{both})$_1$. For the inhomogeneous field
equation, we propose the auxiliary {\it Lagrangian\/} 4-form
\begin{equation}\label{ALag}
\cL=-\frac{1}{2\Omega_0}\left(\cF_\a\wedge\,^\star\!\cF^\a+{}^\star\!
R^{\a\b}\wedge\cA_\a\wedge \cA_\b\right)\,.
\end{equation}
This is the Lagrangian for a massless Lorentz vector valued 1-form
field that is non-minimally coupled to the curvature.  Variation with
respect to $\cA_\a$ yields
\begin{equation}\label{inhA'}
D\,^\star\!\cF^\a={}^\star\!R_\b{}^\a\wedge\cA^\b\,,
\end{equation}
which coincides with (\ref{both})$_2$. Note that the Lagrangian
(\ref{ALag}), if Evans' ansatz is substituted, is similar in structure
as the improved Evans Lagrangian I, Eq.(77). However, (\ref{ALag}) is
a pure electromagnetical Lagrangian whereas I, Eq.(77) is purely
gravitational.

\subsection{Energy-momentum}

Having recovered the (unsubstituted) electromagnetic field equations,
we turn to the energy question. In the EC-theory, we get the
energy-momentum by varying the Lagrangian with respect to the coframe:
\begin{eqnarray}\label{simax}
  \Sigma_\alpha^{\rm elmg} := - \frac{\d\cL}{\d\vt^\a}&=&
  \frac{1}{2\Omega_0}\left[\cF^\b\wedge(e_\alpha\rfloor
    {}^\star\! \cF_\b ) -  {}^\star\! \cF^\b\wedge (e_\alpha\rfloor
    \cF_\b)\right.\\ &&\left.+(\cA_\b\wedge\cA_\g)\wedge
\left(e_\a\rfloor{}^\star\!R^{\b\g}
    \right)
    -\,^\star(\cA_\b\wedge\cA_\g)\wedge(e_\a\rfloor R^{\b\g}) \right]\,.
  \nonumber
\end{eqnarray}
Here we need the master formula of \cite{Muench} for the commutator of
a variation $\d$ with the Hodge star operator $^\star$. The
energy-momentum is still tracefree as in Maxwell's
theory,
\begin{equation}\label{tracefree}
\vt^{\a}\wedge  \Sigma_{\a}^{\rm elmg}=0\,,
\end{equation}
since $\cA^\a$ is a massless field, and, perhaps surprisingly, the
energy-momentum remains symmetric,
\begin{equation}\label{antisymm}
  \vt_{[\a}\wedge  \Sigma_{\b]}^{\rm elmg}
  =\frac{1}{2\Omega_0}\vt_{[\a}\wedge\left[\left(e_{\b]}\rfloor{}^\star\!R^{\g\d}
  \right)\wedge(\cA_\g\wedge\cA_\d)
  -(e_{\b]}\rfloor R^{\g\d})\wedge\,^\star(\cA_\g\wedge\cA_\d) \right]=0,
\end{equation}
as some algebra\footnote{Let $\Phi$ be an arbitrary $p$-form in a
  4-dimensions RC-space with Lorentzian signature. After applying the
  formula $^\star(\Phi\wedge \vt_\a)=e_\a\rfloor\,^\star \Phi$ twice,
  it can be shown that $\vt_{[\a}\wedge
  e_{\b]}\rfloor\,^\star\Phi=\,^\star\!\left(\vt_{[\a}\wedge
    e_{\b]}\rfloor \Phi\right)$.}
shows, compare \cite{Birkbook}, Eqs.(B.5.20) and (E.1.27).

We substitute Evans' ansatz (\ref{Ansatz}) and find
\begin{eqnarray}\nonumber
  \Sigma_\alpha^{\rm elmg}&=&\frac{a_0^2}{2\Omega_0}
\left[T^\b\wedge(e_\alpha\rfloor
    {}^\star T_\b) -  {}^\star T_\b\wedge (e_\alpha\rfloor
    T_\b)\right.\\
  &&+\left.(\vt_\b\wedge\vt_\g)\wedge\left(e_\a\rfloor{}^\star\!R^{\b\g} \right)
    -\,^\star(\vt_\b\wedge\vt_\g)\wedge(e_\a\rfloor R^{\b\g}) \right]
  \,.\label{simax'}
\end{eqnarray}
By some algebra, the term in the second line can be a bit simplified:
\begin{eqnarray}\nonumber
  \Sigma_\alpha^{\rm elmg}&=&\frac{a_0^2}{2\Omega_0}
\left[T^\b\wedge(e_\alpha\rfloor
    {}^\star T_\b) -  {}^\star T^\b\wedge (e_\alpha\rfloor
    T_\b)\right.\\
  &&+\left.2\,^\star\!R_{\b\a}\wedge \vt^\b+R^{\b\g}\wedge\eta_{\a\b\g}
  \right]
  \,.\label{simax''}
\end{eqnarray}
Also after the substitution of Evans' ansatz the energy-momentum
remains traceless $\vt^\a\wedge \Sigma_\alpha^{\rm elmg}=0$ and
symmetric $ \vt_{[\a}\wedge  \Sigma_{\b]}^{\rm elmg} =0$.
 
\subsection{Spin angular momentum}

Now we turn to spin angular momentum. Since the extended
electromagnetic potential $\cA^\a$ transforms as a vector under
Lorentz transformations, it carries spin, as any other Lorentz vector
field. Again we find no help in Evans' work. We vary the Lagrangian
(\ref{ALag}) with respect to the connection:
\begin{eqnarray}
  \tau_{\a\b}^{\rm elmg}:=-\frac{\delta \cL}{\delta\Gamma^{\a\b}}=
  \frac{1}{\Omega_0}\left[\cA_{[\a}\wedge
    {}^\star\! \cF_{\b]}+\frac{1}{2}\,D\,^\star\!\left(\cA_\a\wedge\cA_\b
    \right)\right]\,.
\label{spin}
\end{eqnarray}
If we substitute Evans' ansatz (\ref{Ansatz}), we get
\begin{eqnarray}
  \tau_{\a\b}^{\rm elmg}=
  \frac{a_0^2}{\Omega_0}\left(\vt_{[\a}\wedge
    {}^\star T_{\b]}+\frac{1}{2}\,\eta_{\a\b\g}\wedge T^\g\right)\,.
\label{spin'}
\end{eqnarray}

Now we apply the exterior covariant derivative to (\ref{spin}):
\begin{eqnarray}
  D \tau_{\alpha\b}^{\rm  elmg}=\frac{1}{\Omega_0}\left[(\underbrace{
      \cF_{[\a}\wedge{}^\star\! \cF_{\b]}}_{=0}-\cA_{[\a}\wedge D\,^\star\!
    \cF_{\b]} ) +\frac 12\,DD\,^\star\!\left(\cA_\a\wedge\cA_\b 
    \right)\right]  \label{Dtau}\,.
\end{eqnarray}
After using the inhomogeneous field equation and the Ricci identity,
we find
\begin{eqnarray}\label{Dtau'}
 D \tau_{\alpha\b}^{\rm  elmg}=0\,.
\end{eqnarray}
Thus, the spin of the field $\cA^\a$, without contribution of the
$\cA^\a$-field's orbital angular momentum, is covariantly
conserved.  As we see {}from (\ref{antisymm}) and (\ref{Dtau'}), angular
momentum conservation for the vacuum case is fulfilled:
\begin{equation}
  D \tau_{\alpha\b}^{\rm  elmg}+\vt_{[\a}\wedge  \Sigma_{\b]}^{\rm elmg}=0\,.
\end{equation}

\section{A new variational principle for gravity and extended
  electromagnetism}

Evans' theory is distinguished {}from the foregoing system by the new
ansatz (\ref{Ansatz}) for electromagnetism. Thus, instead of the
Maxwell Lagrangian, as in (\ref{ECMax}), we have to take the new
Lagrangian (\ref{ALag}) describing the Evans field $\cA^\a$. Adding a
Lagrange multiplier piece that enforces Evans' ansatz, we find
\begin{eqnarray}\label{principle}
  L&=&-\frac{1}{2\kappa}\,^\star\!\left(\vt_\a\wedge\vt_\b \right)
  \wedge R^{\a\b}-
  \frac{1}{2\Omega_0}\left(\cF_\a\wedge\,^\star\!\cF^\a+{}^\star\!
    R^{\a\b}\wedge\cA_\a\wedge \cA_\b\right)\nonumber\\&& +L_{\rm
    mat}\left(\vt^\a, \Psi^{\a\b...},D\Psi^{\a\b...}\right)
  +\lambda_\a\wedge\left(\cA^\a-a_0\,\vt^\a \right)\,.
\end{eqnarray}
The Lagrange multiplier is a covector-valued 3-form with 16
independent components. The conserved currents of this model
Lagrangian can be derived with the help of the general formalism as
developed, e.g., by Obukhov and Rubilar \cite{Obukhov2006b}.

Let us first discuss the situation when the Lagrange multiplier is put
to zero. Then variations with respect to $\cA^\a, \vt^\a,\Gamma^{\a\b}$
lead to the field equations (\ref{both})$_2$, (\ref{Sc1*}), (\ref{Sc2*}),
respectively, that is, apart {}from Evans' ansatz, we recover the
relevant field equations in electromagnetism and gravitation as they
are characteristic for Evans' theory. Insofar the variational
principle does what it is supposed to.

Now we relax the multiplier and, accordingly, have a new field
variable $\lambda_\a$.  If we {\it drop the matter fields}, the variation of
the Lagrangian (\ref{principle}) looks now as follows:
\begin{eqnarray}\label{principle1}
  \d L&=&\hspace{5pt} \d \cA_\a\wedge\left[-\frac{1}{2\Omega_0}
    \left(D^\star\cF^\a
      -{}^\star\!R_\b{}^\a\wedge\cA^\b\right) -\lambda^\a\right]\nonumber\\
  &&+\d \vt^\a\wedge\left[\frac{1}{2\kappa}\left(-\,G_\a
      -\kappa\,\Sigma_\a^{\rm elmg}\right)
    +a_0\lambda_\a \right]\nonumber\\
  &&+\d\Gamma^{\a\b}\wedge\left[\frac{1}{2\kappa}\left(-\,C_{\a\b} 
      -\kappa\,\tau_{\a\b}^{\rm elmg}\right) \right]\nonumber\\
  &&+\d\lambda_\a\wedge\left[ \cA^\a-a_0\vt^\a \right]\,.
\end{eqnarray}
Here $G_\alpha := {\frac 12}\eta_{\alpha\beta\gamma}\wedge
R^{\beta\gamma}$ is the Einstein 3-form and and $C_{\alpha\beta} :=
{\frac 12}\eta_{\alpha\beta\gamma}\wedge T^\gamma$ the Cartan 3-form,
as they were defined in I, Eq.(16) and I, Eq.(15), respectively. They
arise also {}from the variation of the Hilbert type Lagrangian with
respect to coframe and connection.  The expressions in the
brackets have to vanish at the extremum of the action. 

The first term yields the value for the multiplier
\begin{eqnarray}\label{multiplier}
  \lambda^\a=-\,\frac{1}{\Omega_0}\left(D^\star\cF^\a
    -{}^\star\!R_\b{}^\a\wedge\cA^\b\right)=-\,\frac{a_0}{\Omega_0}
  \left(D^\star T^\a -{}^\star\!R_\b{}^\a\wedge\vt^\b \right)\,.
\end{eqnarray}
Consequently, the first field equation of gravitation is modified,
\begin{eqnarray}\label{einstein}
G_\a=\kappa\,\Sigma_\a^{\rm elmg}
    + \frac{a_0^2\kappa}{\Omega_0}\left(D^\star T_\a
  -{}^\star\!R_{\b\a}\wedge\vt^\b \right)\,, 
\end{eqnarray}
whereas the second one remains the same, namely,
\begin{eqnarray}\label{cartan}
C_{\a\b}=\kappa\,\tau_{\a\b}^{\rm
      elmg}\,.
\end{eqnarray}
We introduce the dimensionless constant\footnote{\;We determine the
  dimensions of the different pieces ($\ell$ dimension of length,
  $\frak{h}$ of action, $q$ of electric charge, $\phi$ of magnetic
  flux):
  \begin{equation}
    [a_0]=\frac{\Phi}{\ell}=\frac{\frak{h}}{q\ell}\,,\quad
    [\kappa]=\frac{[\kappa]\,\frak{h}}{\frak{h}}=\frac{\ell^2}
    {\frak{h}}\,,
    \quad [\Omega_0]=\frac{\frak{h}}{q^2}\,,\quad\Rightarrow\quad
    [\xi]=\left(\frac{\frak{h}}{q\ell}\right)^2
    \frac{\ell^2}{\frak{h}}
    \frac{q^2}{\frak{h}}=1\,.\nonumber
  \end{equation}}
\begin{equation}\label{dimensionless}
\xi:=\frac{a_0^2\kappa}{\Omega_0}\,,
\end{equation}
which is characteristic for Evans' theory. Using the length $\ell_{\rm
  E}$, see I, Eq.(25), we have $a_0=h/(2e\ell_{\rm E})$. Since
$\kappa=8\pi G/c^3$, we find
\begin{equation}
  \xi= \frac{h^2}{4e^2\ell_{\rm E}^2}\, \frac{8\pi G}{c^3\Omega_0} 
 =2\pi^2\frac{2\Omega_{\rm QHE}}{\Omega_0}\left(\frac{\ell_{\rm P}}
{\ell_{\rm E}}\right)^2=\frac{2\pi^2}{\a}\,\left(\frac{\ell_{\rm P}}
{\ell_{\rm E}}\right)^2\approx 2705\left(\frac{\ell_{\rm P}}
{\ell_{\rm E}}\right)^2\,,
\end{equation}
with $\ell_{\rm P}:=\sqrt{G\hbar/c^3}\approx 10^{-31}m$ as Planck
length, $\Omega_{\rm QHE}:=h/e^2$ as Quantum Hall resistance (von
Klitzing constant), see \cite{Flowers}, and $\a$ as fine structure
constant. Note, since $G>0$ and $\Omega_0>0$, we have always $\xi>0$.

All what is left {}from Evans' theory, are these 16+24 field equations
in which the sources are specified by (\ref{simax''}) and
(\ref{spin'}), but only 10 of the 16 are independent. If we substitute
these sources, we find the system
\begin{eqnarray}\label{einstein'}
-\,\frac 12\,\eta_{\a\b\g}\wedge R^{\b\g}&=&
\frac{\xi}{2}
\Big[T^\b\wedge(e_\alpha\rfloor
    {}^\star T_\b) -  {}^\star T_\b\wedge (e_\alpha\rfloor
    T_\b)\nonumber\\
&& + R^{\b\g}\wedge\eta_{\a\b\g} + 2D\,{}^\star T_\a\Big]\,,\\ 
\label{cartan'}
-\,\frac 12\,\eta_{\a\b\g}\wedge T^{\g}&=&\xi\left(\vt_{[\a}\wedge
    {}^\star T_{\b]}+\frac{1}{2}\,\eta_{\a\b\g}\wedge T^\g\right)\,.
\end{eqnarray}
Note that (\ref{einstein'}) represents partial differential equations
of second order in the coframe, because $D\,^\star
T_\a=D\,^\star\! D\vt_\a$, and first order in the connection. The
linearized version is a wave type equation for the coframe $\vt_\a$.
\section{Solution of the second field equation}

The second (Cartan's) field equation (\ref{cartan'}) is a homogeneous
algebraic equation for the components of the torsion. We can solve
this equation exactly. For this purpose we need the identity (for the proof,
see \cite{Obukhov2006a}, for example):
\begin{eqnarray}
\frac 12\,\eta_{\a\b\g}\wedge T^{\g} &\equiv& \vartheta_{[\alpha}\wedge
h_{\beta]},\label{ID1}\\
h_\alpha &:=& {}^\star(-{}^{(1)} T_\alpha + 2{}^{(2)}T_\alpha 
+ {\frac 1 2}{}^{(3)}T_\alpha).\label{ID2}
\end{eqnarray}
The right-hand side of (\ref{ID2}) is constructed {}from the irreducible 
parts of the torsion. Namely, let us recall that the torsion 2-form can 
be decomposed into the three irreducible pieces, 
\begin{equation}
T^{\alpha}={}^{(1)}T^{\alpha} + {}^{(2)}T^{\alpha} + 
{}^{(3)}T^{\alpha},\label{TorI}
\end{equation}
where the {\it vector}, {\it axial vector} and {\it pure tensor}
parts of the torsion are defined by 
\begin{eqnarray}
{}^{(2)}T^{\alpha}&=& {1\over 3}\vartheta^{\alpha}\wedge (e_\nu\rfloor 
T^\nu),\label{iT2}\\
{}^{(3)}T^{\alpha}&=& -\,{1\over 3}{}^\star(\vartheta^{\alpha}\wedge{}^\star
(T^{\nu}\wedge\vartheta_{\nu}))= {1\over 3}e^\alpha\rfloor(T^{\nu}\wedge
\vartheta_{\nu}),\label{iT3}\\
{}^{(1)}T^{\alpha}&=& T^{\alpha}-{}^{(2)}T^{\alpha} - {}^{(3)}T^{\alpha}.
\label{iT1}
\end{eqnarray}

Substituting (\ref{ID1}) into (\ref{cartan'}), we find
\begin{equation}
\xi\,{}^\star T_\alpha + (1 + \xi)\,h_\alpha = 0. 
\end{equation}
Using then (\ref{ID2}) and (\ref{TorI}), we can ultimately recast the
last equation into the form
\begin{equation}
-\,{}^{(1)}T^{\alpha} + (3\xi + 2)\,{}^{(2)}T^{\alpha} + 
{\frac 12}(3\xi + 1)\,{}^{(3)}T^{\alpha} = 0.\label{T123}
\end{equation}
The irreducible parts are all algebraically independent. Hence we can
conclude that all the three terms in (\ref{T123}) vanish. For generic
case of the coupling constant $\xi$ we thus ultimately find the
trivial solution:
\begin{equation}
{}^{(1)}T^{\alpha} = {}^{(2)}T^{\alpha} = {}^{(3)}T^{\alpha} = 0,\quad
{\rm hence}\quad T_\alpha = 0.\label{zeroT}
\end{equation}
We may have nontrivial torsion for two exceptional cases. Namely, when
\begin{equation}\label{xi1}
\xi = -\,{\frac 23}\qquad\mbox{or}\qquad \xi = -\,{\frac 13}\,.
\end{equation}
However,  since  $\xi>0$,  these  are  unphysical cases  that  can  be
excluded.

\section{Conclusions}

For the generic case, we substitute the vanishing torsion solution 
(\ref{zeroT}) into the first field equation (\ref{einstein'}) and 
find that the latter reduces to the usual Einstein equation 
\begin{equation}\label{einstein''}
\frac 12\,\eta_{\a\b\g}\wedge \widetilde{R}^{\b\g} = 0,
\end{equation}
where the tilde denotes the object constructed {}from the Riemannian
(Christoffel) connection. In this sense, the model under consideration
is similar to the Einstein-Cartan theory that also reduces to
Einstein's general relativity in absence of the sources with spin, see
\cite{RMP,Trautman}.

This similarity goes even further when the nontrivial matter sources
are taken into account. Then the right-hand side of (\ref{cartan'}) will
contain the spin current 3-form $\tau^{\rm mat}_{\alpha\beta}$ of the
matter fields. Subsequently one can solve the second field 
equation (\ref{cartan'}), expressing the torsion in terms of the 
spin of matter. By substituting this into (\ref{einstein'}), we can 
recast the first field equation into a form of the Einstein equation
with the {\it effective} energy-momentum current that will contain the 
quadratic contributions of spin. The same occurs in the Einstein-Cartan
theory, too. 

\subsubsection*{Acknowledgments} We are grateful to Robert G.~Flower
(Applied Science Associates) for helpful remarks. This work has been
supported by the grant HE 528/21-1 of the DFG (Bonn).

\begin{footnotesize}

\end{footnotesize}
\end{document}